\newcommand*{\ATLASLATEXPATH}{latex/}
\documentclass[cernpreprint,texlive=2016,UKenglish]{\ATLASLATEXPATH atlasdoc}
\usepackage[block=space, maxbibnames=5]{\ATLASLATEXPATH atlaspackage}
\usepackage[articletitle=true]{\ATLASLATEXPATH atlasbiblatex}
\usepackage[process=true]{\ATLASLATEXPATH atlasphysics}
\graphicspath{{logos/}{figs/}}
\usepackage{TOPQ_2015_16-defs}
\AtlasTitle{Measurement of the cross-section for producing a $W$~boson in
  association with a single top quark in $pp$ collisions at $\sqs=\SI{13}{\TeV}$
    with ATLAS}

\author{The ATLAS Collaboration}
\AtlasRefCode{TOPQ-2015-16}
\AtlasNote{ATL-COM-PHYS-2016-874}
\AtlasJournal{JHEP}
\AtlasJournalRef{JHEP 01 (2018) 63}
\AtlasDOI{DOI:10.1007/JHEP01(2018)063}
  \AtlasAbstract{%
    The inclusive cross-section for the associated production of a $W$ boson and top quark is measured using data from proton--proton collisions at $\sqs=\SI{13}{\TeV}$. The dataset corresponds to an integrated luminosity of \lumi, and was collected in 2015 by the ATLAS detector at the Large Hadron Collider at CERN.
    Events are selected requiring two opposite sign isolated leptons and at least one jet; they are
    separated into signal and control regions based on their jet multiplicity and the number of jets that are identified as containing $b$ hadrons.
    The $Wt$ signal is then separated from the \ttbar background using boosted decision tree discriminants in two regions.
    The cross-section is extracted by fitting templates to the data distributions, and is measured to be \fullXsSplit{}.
The measured value is in good agreement with the SM prediction of \ensuremath{\sigma_{\text{theory}}=\xspace{\SI[parse-numbers=false]{71.7\pm1.8\,(\mathrm{scale})\pm{3.4}\,(\mathrm{PDF})}{\fullXsUnit}}}~\cite{Kidonakis:2015}.
  }

\hypersetup{pdftitle={ATLAS draft},pdfauthor={The ATLAS Collaboration}}
\begin{document}
\maketitle
\tableofcontents
\section{Introduction}
\label{sec:intro}
\begin{figure}[ht]
  \centering
  \vspace{12pt}
  \begin{fmffile}{tchan1}
    \begin{fmfgraph*}(100,60)
      \fmfset{arrow_len}{3mm}
      \fmfstraight
      \fmfleft{i1,i2}
      \fmfright{o1,o2}
      \fmflabel{$q$}{i1}
      \fmflabel{$b$}{i2}
      \fmflabel{$q'$}{o1}
      \fmflabel{$t$}{o2}
      \fmf{fermion,tension=1}{i1,v1}
      \fmf{fermion,tension=1}{i2,v2}
      \fmf{boson,label=$W$,tension=1}{v1,v2}
      \fmf{fermion,tension=1}{v1,o1}
      \fmf{fermion,tension=1}{v2,o2}
    \end{fmfgraph*}
  \end{fmffile}
  \hspace{4ex}
  \begin{fmffile}{tchan2}
    \begin{fmfgraph*}(100,55)
      \fmfset{arrow_len}{3mm}
      \fmfstraight
      \fmfleft{i1,i3,i2}
      \fmfright{o1,o2,o3}
      \fmflabel{$q$}{i1}
      \fmflabel{$g$}{i2}
      \fmflabel{$q'$}{o1}
      \fmflabel{$t$}{o2}
      \fmflabel{$\bar{b}$}{o3}
      \fmf{gluon,tension=1}{i2,v3}
      \fmf{phantom,tension=1}{i3,v2}
      \fmf{fermion,tension=0}{v3,v2}
      \fmf{fermion,tension=1}{v2,o2}
      \fmf{fermion,tension=1}{o3,v3}
      \fmffreeze
      \fmf{fermion,tension=1}{i1,v1}
      \fmf{boson,label=$W$,tension=1,label.side=left}{v1,v2}
      \fmf{fermion,tension=1}{v1,o1}
    \end{fmfgraph*}
  \end{fmffile}
  \hspace{2cm}
  \begin{fmffile}{schan}
    \begin{fmfgraph*}(100,55)
      \fmfset{arrow_len}{3mm}
      \fmfstraight
      \fmfleft{i1,i2}
      \fmfright{o1,o2}
      \fmflabel{$q$}{i1}
      \fmflabel{$\bar{q}'$}{i2}
      \fmflabel{$\bar{b}$}{o1}
      \fmflabel{$t$}{o2}
      \fmf{fermion,tension=1}{i1,v1,i2}
      \fmf{boson,label=$W$,tension=1}{v1,v2}
      \fmf{fermion,tension=1}{o1,v2,o2}
    \end{fmfgraph*}
  \end{fmffile}
  \\[2em]\raggedright \hspace{4.8cm}(a) \hspace{7.8cm}(b)
  \setlength{\abovecaptionskip}{10pt plus 6pt minus 3pt}
  \caption{%
    Representative leading-order Feynman diagrams for the production of a single top quark in (a) the \hyNounChan{$t$}-channel and (b) the \hyNounChan{$s$}-channel.%
}
  \label{fig:feyn_other}
\end{figure}
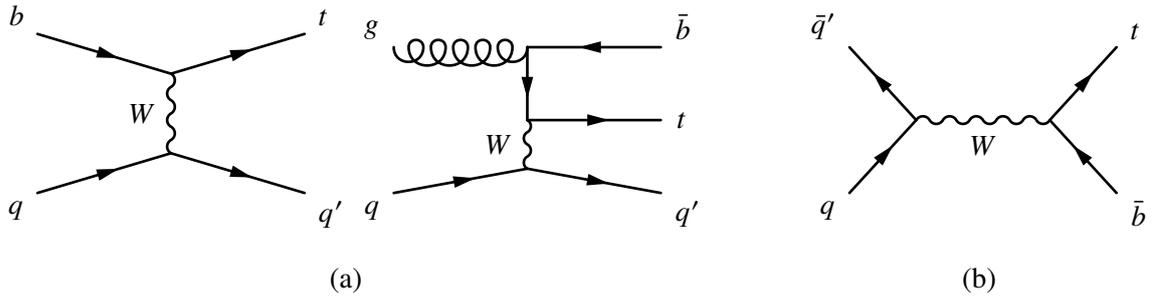

Top quarks can be produced singly via electroweak interactions involving a $Wtb$ vertex.
In the Standard Model (SM), single top quark production proceeds via three channels at leading order (LO), represented in \Figs{\ref{fig:feyn_wt}}{\ref{fig:feyn_other}}:
production in association with a $W$ boson ($Wt$),
the $t$-channel and
the $s$-channel.
At the Large Hadron Collider (LHC), the \hyNounChan{$Wt$}channel is the mode with the second largest production cross-section, behind the dominant $t$-channel mode.
The \hyNounChan{$Wt$}channel represents approximately \SI{24}{\%} of the total single-top-quark production rate at \sqs$=\SI{13}{\TeV}$, making it experimentally accessible for detailed measurements.

\begin{figure}[h]
  \centering
  \vspace{12pt}
  \begin{fmffile}{wt}
    \begin{fmfgraph*}(150,80)
      \fmfset{arrow_len}{3mm}
      \fmfstraight
      \fmfleft{i1,i2}
      \fmfright{o1,o2,o3,o4,o5}
      \fmflabel{$g$}{i1}
      \fmflabel{$b$}{i2}
      \fmflabel{$\ell^{+}$}{o1}
      \fmflabel{$\nu$}{o2}
      \fmflabel{$b$}{o3}
      \fmflabel{$\ell^{-}$}{o4}
      \fmflabel{$\bar{\nu}$}{o5}
      \fmf{gluon,tension=2}{i1,v1}
      \fmf{fermion,tension=2}{i2,v1}
      \fmf{fermion,label=$b$,tension=2.5}{v1,v2}
      \fmf{fermion,label=$t$,tension=0.8}{v2,v3}
      \fmf{fermion,tension=0.5}{v3,o3}
      \fmf{boson,label=$W^{+}$,tension=0.2}{v3,v4}
      \fmf{fermion,tension=0.2}{o1,v4,o2}
      \fmf{boson,label=$W^{-}$,label.side=left}{v2,v5}
      \fmf{fermion,tension=0.5}{o5,v5,o4}
    \end{fmfgraph*}
  \end{fmffile}
  \setlength{\abovecaptionskip}{22pt plus 6pt minus 3pt}
  \caption{%
    A representative leading-order Feynman diagram for the production of a single top quark in the \hyNounChan{$Wt$}channel and the subsequent leptonic decay of both the $W$ boson and top quark.%
}
  \label{fig:feyn_wt}
\end{figure}
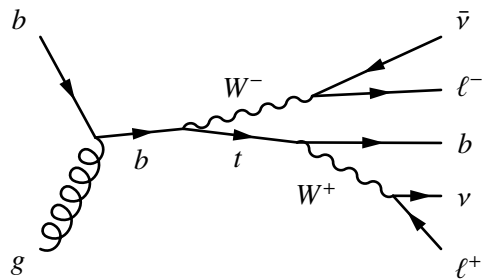

The cross-section for each of the three single-top-quark production channels is sensitive to the coupling between the $W$ boson and the top quark.
This coupling is parameterised by the relevant Cabibbo--Kobayashi--Maskawa (CKM) matrix element $V_{tb}$ and form factor $f^\mathrm{L}_\mathrm{V}$~\cite{coupling_params1,coupling_params2,coupling_params3} such that the proportionality is given by $|f^\mathrm{L}_\mathrm{V} V_{tb}|^{2}$~\cite{coupling_prop1,coupling_prop2}, assuming a left-handed vector interaction as given in the SM.
Single top quark production therefore presents an opportunity for testing the structure of the SM, as well as probing classes of new-physics models that can affect the $Wtb$ vertex.
In contrast to the $t$- and $s$-channels, which are sensitive to both the existence of four-fermion operators and corrections to the $Wtb$ vertex, the \hyNounChan{$Wt$}channel only depends on the latter; it is therefore important to study this channel separately to provide a comparison with the other channels~\cite{motiv1,motiv2}.

The \hyNounChan{$Wt$}channel was not accessible at the Tevatron due to its small cross-section in $p\bar{p}$ collisions at $\sqs= \SI{1.96}{\TeV}$.
At the LHC, however, evidence of this process with \SI{7}{\TeV} collision data was presented by the ATLAS Collaboration~\cite{TOPQ-2011-17}
and by the CMS Collaboration~\cite{CMS-TOP-11-022}. 
With \SI{8}{\TeV} collision data, observations were made by the CMS Collaboration~\cite{CMS-TOP-12-040} and the ATLAS Collaboration~\cite{TOPQ-2012-20} with cross-section measurements in good agreement with theoretical predictions.

The $Wt$ NLO cross-section at a $\sqs=\SI{13}{\TeV}$ with next-to-next-to-leading logarithmic (NNLL) soft-gluon corrections is calculated as
\ensuremath{\sigma_{\text{theory}}=\xspace{\SI[parse-numbers=false]{71.7\pm1.8\,(\mathrm{scale})\pm{3.4}\,(\mathrm{PDF})}{\fullXsUnit}}}~\cite{Kidonakis:2015},
assuming a top quark mass ($m_{\text{top}}$) of \SI{172.5}{\GeV}.
The first uncertainty accounts for the renormalisation and factorisation scale variations (from $m_{\text{top}}/2$ to $2 \,m_{\text{top}}$), while the second uncertainty originates from uncertainties in the MSTW2008 NLO parton distribution function (PDF) sets~\cite{mstw}.

This \doctype describes a measurement of the cross-section of the $Wt$ process using
$\sqs = \SI{13}{\TeV}$ proton--proton ($pp$) collisions with an integrated luminosity of \lumi. The data were recorded with the ATLAS detector in 2015.
The measurement is made using events containing at least one $b$ jet (according to the definition in \Sect{\ref{sec:objects}}) and exactly two oppositely charged leptons in the final state, where a lepton ($\ell$) is defined to be either an electron ($e$) or a muon ($\mu$), whether produced directly from the decay of a $W$ boson or from the decay of an intermediate $\tau$ lepton.
The $Wt$ signal enters this final state when the top quark decays into a $W$ boson and a quark (which is assumed to be a $b$-quark), with both $W$ bosons subsequently decaying into a neutrino and a lepton, as depicted in \Fig{\ref{fig:feyn_wt}}.
A minimal selection is applied to reduce background contributions from $Z/\gamma^{*}+$jets (hereafter called \Zjets) events, diboson events, and events containing leptons that are misidentified or arise from the decay of hadrons.
A boosted decision tree (BDT) analysis is performed to construct discriminants capable of separating the $Wt$ signal from the dominant top quark pair (\ttbar{}) background, and these discriminants are used in a profile-likelihood fit to extract the $Wt$ cross-section.
The top pair production background is described by simulation, which has been validated in previous ATLAS measurements~\cite{TOPQ-2015-09}.

The measurement technique is similar to that employed in the corresponding \SI{8}{\TeV} ATLAS measurement~\cite{TOPQ-2012-20}.
The most significant changes include modifications to the BDT training and the binning of the distribution used in the likelihood fit (discussed in \Sect{\ref{sec:mva}} and \Sect{\ref{sec:fit}} respectively), and an optimisation of kinematic requirements to more effectively reject \Zjets{} and other small backgrounds (discussed in \Sect{\ref{sec:events}}).

\section{The ATLAS detector}
\label{sec:detector}
The ATLAS detector~\cite{PERF-2007-01} at the LHC covers nearly the entire solid angle%
\footnote{%
ATLAS uses a right-handed coordinate system with its origin at the nominal interaction point (IP) in the centre of the detector and the $z$-axis along the beam pipe. The $x$-axis points from the IP to the centre of the LHC ring, and the $y$-axis points upward. Cylindrical coordinates $(r,\phi)$ are used in the transverse plane, $\phi$ being the azimuthal angle around the $z$-axis. The pseudorapidity is defined in terms of the polar angle $\theta$ as $\eta=-\ln\tan(\theta/2)$, while the rapidity is defined in terms of particle energies and the $z$-component of particle momenta as $y=(1/2)\ln\left[(E+p_z)/(E-p_z)\right]$.%
}
around the collision point, and consists of an inner tracking detector (ID) surrounded by a thin superconducting solenoid magnet producing a \SI{2}{T} axial magnetic field, electromagnetic (EM) and hadronic calorimeters, and an external muon spectrometer (MS).
The ID consists of a high-granularity silicon pixel detector and a silicon microstrip tracker, together providing precision tracking in the pseudorapidity
range $|\eta|<2.5$, complemented by a transition radiation tracker providing tracking and electron identification information for $|\eta|<2.0$.
The innermost pixel layer, the insertable B-layer, was added between \RunOne and \RunTwo of the LHC, at an innermost radius of \SI{33}{mm} around a new, thinner, beam pipe~\cite{ATL-PHYS-PUB-2015-051}.
A lead liquid-argon (LAr) electromagnetic calorimeter covers the region $|\eta|<3.2$, and hadronic calorimetry is provided by steel/scintillator tile calorimeters within $|\eta|<1.7$ and copper/LAr hadronic endcap calorimeters in the range $1.5 < |\eta|< 3.2$.
A LAr forward calorimeter with copper and tungsten absorbers covers the range $3.1 < |\eta| <4.9$.
The MS consists of precision tracking chambers covering the region $|\eta|<2.7$, and separate trigger chambers covering $|\eta|<2.4$.
A two-level trigger system, using a custom hardware level followed by a software-based level, selects from the \SI{40}{MHz} of collisions a maximum of around \SI{1}{kHz} of interesting events for offline storage.

\section{Data and simulation}
\label{sec:samples}
The data events analysed in this \doctype correspond to an integrated luminosity of \lumi collected from the operation of the LHC in 2015 at $\sqs=\SI{13}{\TeV}$ with a bunch spacing of \SI{25}{ns} and an average number of collisions per bunch crossing $\langle\mu\rangle$ of around 14.
They are required to be recorded in periods where all detector systems are flagged as operating normally.
Additionally, individual events identified as containing corrupted data are rejected.

Monte Carlo (MC) simulation samples are used to estimate the efficiency to select signal and background events, train and test BDTs, estimate systematic uncertainties, and validate the analysis tools.
All simulation samples are normalised to theoretical cross-section predictions.
The nominal samples (used for estimating the central values for efficiencies and background templates) were simulated with a full ATLAS detector simulation~\cite{SOFT-2010-01} implemented in \GEANT~4~\cite{geant4}.
Many of the samples used in the estimation of systematic uncertainties were instead produced using \textsc{Atlfast2}~\cite{ATL-PHYS-PUB-2010-013}, which differs from the full simulation in that the ATLAS calorimeters and their responses are simulated using a faster approximation.
Pile-up (additional $pp$ collisions in the same or a nearby bunch crossing) was included in the simulation by overlaying collisions with the soft QCD processes of \PYTHIA~8.186~\cite{pythia81} using a set of tuned parameters called the  A2 tune~\cite{ATL-PHYS-PUB-2012-003} and the MSTW2008LO PDF set.
Events were generated with a predefined distribution of the expected number of interactions per bunch crossing, then reweighted to match the actual observed data conditions.
In all samples used for this analysis $m_{\text{top}}$ was set to \SI{172.5}{\GeV} and the $W\to \ell \nu$ branching ratio was set to $0.1080$ per lepton flavour.

For the generation of $Wt$ and \ttbar{} event samples~\cite{ATL-PHYS-PUB-2016-004}, the \POWHEGBOX v1 (v2 for \ttbar)~\cite{Nason:2004rx,Frixione:2007vw,Alioli:2010xd,Re:2010bp,Campbell:2014kua} generator with the CT10 PDF set~\cite{ct10} in the matrix element calculations is used.
For these processes, top quark spin correlations are preserved.
The parton shower, fragmentation, and underlying event were simulated using \PYTHIAV{6.428}~\cite{pythia6} with the CTEQ6L1 PDF set~\cite{cteq6} and the corresponding Perugia~2012 (P2012) tune~\cite{perugia}.
The \textsc{EvtGen}~v1.2.0 program~\cite{evtgen} was used to simulate properties of the bottom and charmed hadron decays.
The renormalisation and factorisation scales are set to $m_{\text{top}}$ for the $Wt$ process and $\sqrt{m_{\text{top}}^2 + \pt(t)^2}$ for the \ttbar process.
The diagram removal (DR) scheme~\cite{dsdr}, in which all next-to-leading order (NLO) diagrams that overlap with the \ttbar\ definition are removed from the calculation of the $Wt$ amplitude, was employed to handle interference between $Wt$ and \ttbar\ diagrams, and was applied to the $Wt$ sample.

The \ttbar cross-section is set to $\sigma_{\ttbar} = \SI[parse-numbers=false]{252.9\,^{+6.4}_{-8.6}\,(\mathrm{scale})\,\pm11.7\,(\mathrm{PDF} + \alphas)}{pb}$ as calculated with the \texttt{Top++2.0} program to NNLO, including soft-gluon resummation to NNLL~\cite{top++}.
The first uncertainty comes from the independent variation of the factorisation and renormalisation scales, $\mu_{\mathrm{F}}$ and $\mu_{\mathrm{R}}$, while the second one is associated with variations in the PDF and $\alphas$, following the PDF4LHC prescription with the MSTW2008 \SI{68}{\%} CL NNLO, CT10 NNLO and NNPDF2.3 5f FFN PDF set~\cite{Botje:2011sn,Martin:2009bu,Gao:2013xoa,Ball:2012cx}.
Both calculations assume $m_{\text{top}} = \SI{172.5}{\GeV}$.

Additional $Wt$ samples were generated to estimate major systematic uncertainties.
An alternative $Wt$ sample was generated using the diagram subtraction (DS) scheme instead of DR, where a gauge-invariant subtraction term modifies the NLO $Wt$ cross-section to locally cancel the double-resonant \ttbar\ contribution~\cite{dsdr}.
Another sample generated with \MGMCatNLO~\cite{mg5amc2} (instead of the \POWHEGBOX) interfaced with \HERWIGpp~2.7.1~\cite{herwig} using \textsc{Atlfast2} fast simulation is used to estimate uncertainties associated with the modelling of the NLO matrix element generator. A sample generated with \POWHEGBOX interfaced with \HERWIGpp (instead of \PYTHIAV{6}) is used to estimate uncertainties associated with the parton shower, hadronisation, and underlying-event models.
In both cases the UE-EE-5 tune of~\Ref{\cite{ATL-PHYS-PUB-2014-021}} was used for the underlying event, and \textsc{EvtGen}~v1.2.0 was used to simulate properties of the bottom and charmed hadron decays.
Finally, in order to estimate uncertainties arising from additional QCD radiation in the $Wt$ events, a pair of samples were generated with \POWHEGBOX interfaced with \PYTHIAV{6} using \textsc{Atlfast2} and the P2012 tune with higher and lower radiation relative to the nominal set, together with varied renormalisation and factorisation scales.
In these samples the resummation damping factor was doubled in the case of higher radiation.
In order to avoid comparing two different detector response models when estimating systematic uncertainties, another version of the nominal \POWHEGBOX with \PYTHIAV{6} sample was also produced with fast simulation.

Additional \ttbar samples were also generated to estimate major systematic uncertainties.
As with the additional $Wt$ samples, these are used to estimate the uncertainties associated with the matrix element generator (a sample produced using \textsc{Atlfast2} fast simulation with \MGMCatNLO interfaced with \HERWIGpp~2.7.1), parton shower and hadronisation models (a sample produced using \textsc{Atlfast2} with \POWHEGBOX interfaced with \HERWIGpp~2.7.1) and additional QCD radiation (a pair of samples produced using full simulation with the P2012 higher and lower radiation-varied sets of parameters, as well as with varied renormalisation and factorisation scales).

Samples used to model the \Zjets background~\cite{ATL-PHYS-PUB-2016-003} were simulated with \SHERPAV{2.1.1}~\cite{sherpa}.
Matrix elements were calculated for up to two partons at NLO and four partons at LO using the \textsc{Comix}~\cite{comix} and \textsc{OpenLoops}~\cite{openloops} matrix element generators and merged with the \SHERPA parton shower~\cite{csshower} using the \textsc{ME+PS@NLO} prescription~\cite{mepsnlo}.
The CT10 PDF set was used in conjunction with \SHERPA parton shower tuning, with a generator-level cutoff on the dilepton invariant mass of $\mll>\SI{40}{\GeV}$ applied.
The \Zjets events are normalised to NNLO cross-sections.

Diboson processes with four charged leptons, three charged leptons and one neutrino, or two charged leptons and two neutrinos~\cite{ATL-PHYS-PUB-2016-002} were simulated using the \SHERPAV{2.1.1} generator.
The matrix elements contain all diagrams with four electroweak vertices.
The NLO corrections are used for the purely leptonic final states as well as for final states with two or four charged leptons plus one additional parton.
For other final states with up to three additional partons, the LO calculations of the \textsc{Comix} and \textsc{OpenLoops} generators are used.
Their outputs are combined with the \SHERPA parton shower using the ME+PS@NLO prescription~\cite{mepsnlo}.
The PDF set used was CT10 with dedicated parton shower tuning.
The generator-calculated cross-sections are used for diboson processes (already at NLO).

Finally, the very small \Wjets contribution was simulated using \POWHEGBOX v2 interfaced to the \PYTHIAV{8.186}~\cite{pythia81} parton shower model.
The CT10 PDF set was used in the matrix element.
The AZNLO~\cite{aznlo} tune was used, with PDF set CTEQ6L1, for the modelling of non-perturbative effects, and
the \textsc{EvtGen}~v1.2.0 program was used to simulate properties of the bottom and charmed hadron decays.

\section{Object selection}
\label{sec:objects}
Electron candidates are reconstructed from energy deposits in the EM calorimeter associated with ID tracks~\cite{ATLAS-CONF-2016-024}.
The deposits are required to be in the $|\eta|<2.47$ region, with the transition region between the barrel and endcap EM calorimeters, $1.37<|\eta|<1.52$, excluded.
The candidate electrons are required to have transverse energy $\pT>\SI{20}{\GeV}$.
Further requirements on the electromagnetic shower shape, calorimeter energy to tracker momentum ratio, and other discriminating variables are combined into a likelihood-based object quality selection~\cite{ATLAS-CONF-2016-024}, optimised for strong background rejection.
Candidate electrons also must satisfy requirements on the distance of the ID track to the reconstructed primary vertex in the event, which is identified as the vertex with the largest summed $\pt^2$ of associated tracks.
The transverse impact parameter significance must satisfy $|d_0|/\sigma_{d_{0}} < 5$, and the longitudinal impact parameter must satisfy $|\Delta z_0 \sin \theta |<\SI{0.5}{mm}$.
Electrons are further required to be isolated based on ID tracks and topological clusters in the calorimeter~\cite{PERF-2014-07}, with an isolation efficiency of $\SI[parse-numbers=false]{90(99)}{\%}$ for $\pT=\SI[parse-numbers=false]{25(60)}{\GeV}$.

Muon candidates are identified by matching MS segments with ID tracks~\cite{PERF-2015-10}.
The candidates must satisfy requirements on hits in the MS and on the compatibility between ID and MS momentum measurements to remove fake muon signatures.
Furthermore, they must have $\pT>\SI{20}{\GeV}$ as well as $|\eta|<2.5$ to ensure they are within coverage of the ID.
Candidate muons must satisfy the following requirements on the distance of the combined ID and MS track to the primary vertex: the transverse impact parameter significance must satisfy $|d_0|/\sigma_{d_{0}} < 3$, and the longitudinal impact parameter must satisfy $|\Delta z_0 \sin \theta |<\SI{0.5}{mm}$.
An isolation requirement is imposed based on ID tracks and topological clusters in the calorimeter, and results in an isolation efficiency of $\SI[parse-numbers=false]{90(99)}{\%}$ for $\pT=\SI[parse-numbers=false]{25(60)}{\GeV}$.

Single-lepton triggers used in this analysis are designed to select events containing a high-\pt, well-identified charged lepton~\cite{ATL-DAQ-PUB-2016-001}.
They require a \pt\ of at least \SI{20}{\GeV} for muons and \SI{24}{\GeV} for electrons, and also have requirements on the lepton quality and isolation.
These are complemented by triggers with higher \pt\ thresholds and relaxed isolation and identification requirements to ensure maximum efficiency at higher lepton \pt.

Jets are reconstructed from topological clusters in the calorimeter~\cite{ATL-PHYS-PUB-2015-036} using the anti-$k_t$ algorithm~\cite{antikt1,antikt2} with a radius parameter of \SI{0.4}{}.
They are energy-corrected to account for pile-up and calibrated using a \pt- and $\eta$-dependent correction derived from simulation~\cite{ATL-PHYS-PUB-2015-015}.
They are required to have $\pt>\SI{25}{\GeV}$ and $|\eta|<2.5$.
To suppress pile-up, a discriminant called the jet-vertex-tagger (JVT) is constructed using a two-dimensional likelihood method~\cite{PERF-2014-03}.
For jets with $\pT<\SI{60}{\GeV}$ and $|\eta| < 2.4$ a JVT requirement corresponding to a \SI{92}{\%} efficiency while rejecting \SI{98}{\%} of jets from pileup and noise is imposed.

Jets containing $b$-hadrons ($b$-jets) are tagged using a multivariate discriminant which exploits the long lifetime and large invariant mass of $b$-hadron decay products relative to $c$-hadrons and unstable light hadrons~\cite{PERF-2012-04}.
The discriminant is calibrated to achieve a \SI{77}{\%} \btag efficiency and rejection factor of about 4.5 against jets containing charm quarks ($c$-jet) and 140 against light-quark and gluon jets in a sample of simulated \ttbar\ events~\cite{ATL-PHYS-PUB-2015-022}.
The \btag efficiency in simulation is corrected to the efficiency in data~\cite{ATL-PHYS-PUB-2015-039}.

The missing transverse momentum vector is calculated as the negative vectorial sum transverse momenta of particles in the event.
Its magnitude \met is a measure of the transverse momentum imbalance, primarily due to neutrinos that escape detection.
Energy deposits in the calorimeters are uniquely assigned in order of priority to electrons, jets, and muons found in the event, thus avoiding double counting of signals.
This approach also obviates the need for further overlap removal in the \met calculation, since a single energy deposit cannot be re-assigned to two nearby reconstructed signals.
In addition to the identified electrons, jets and muons, a track-based soft term is included in the \met calculation by considering tracks associated with the hard-scattering vertex in the event but not with an identified electron, jet, or muon~\cite{ATL-PHYS-PUB-2015-023,ATL-PHYS-PUB-2015-027}.

To avoid cases where the detector response to a single physical object is reconstructed as two separate final-state objects, several steps are followed to remove such overlaps.
Bremsstrahlung radiation by a muon can result in ID tracks and a calorimeter energy deposit that are also reconstructed as an electron candidate.
Therefore in cases where an electron and muon candidate share an ID track, the object is considered to be a muon, and the electron candidate is rejected.

The overlap of objects is measured using the Lorentz-invariant distance $\Delta R_{y,\phi} = \sqrt{(\Delta y)^{2} + (\Delta\phi)^{2}}$.
Due to the isolation requirements placed on electron candidates, any jets that closely overlap an electron candidate within a cone $\Delta R_{y,\phi} < 0.2$ are likely to be reconstructions of the electron and so are rejected.
When jets and electrons are found within the larger hollow cone $0.2 < \Delta R_{y,\phi} < 0.4$, it is more likely that a real hadronic jet is present and that the electron is a non-prompt constituent of the jet arising from the decay of heavy-flavour hadrons.
Hence and electron candidates found within a cone $\Delta R_{y,\phi} < 0.4$ of any remaining jet is rejected.

Muons can be accompanied by a hard photon due to bremsstrahlung or collinear final state radiation, and the muon-photon system can then be reconstructed as both a jet and muon candidate.
Non-prompt muons can arise from the decay of hadronic jets, however these muons are associated with a higher ID track multiplicity than those accompanied by hard photons.
In order to resolve these ambiguities between nearby jet and muon candidates, first any jets having fewer than three ID tracks and within a cone $\Delta R_{y,\phi} < 0.4$ of any muon candidate are rejected, then any muon candidates within a cone $\Delta R_{y,\phi} < 0.4$ of any remaining jet is rejected.

\section{Event selection and background estimation}
\label{sec:events}
Events are required to have at least one well-reconstructed interaction vertex, to pass a single-electron or single-muon trigger, and to contain at least one jet with $\pt>\SI{25}{\GeV}$.
Events are required to contain exactly two charged leptons of opposite charge with $\pt>\SI{20}{\GeV}$; events with a third lepton with $\pt>\SI{20}{\GeV}$ are rejected.
At least one lepton must have $\pt>\SI{25}{\GeV}$, and at least one of the selected electrons (muons) must be matched within a $\Delta R = \sqrt{(\Delta \eta)^2 + (\Delta\phi)^2}$ cone of size $0.07$ ($0.1$) to the electron (muon) selected online by the corresponding trigger.

In simulated events, information recorded by the event generator is used to identify events in which any selected lepton does not originate promptly from the hard-scatter process.
These non-prompt or fake leptons arise from processes such as the decay of a $b$-hadron, photon conversion or hadron misidentification, and are identified when the electron or muon does not originate from the decay of a $W$ or $Z$ boson (or a $\tau$ lepton itself originating from a $W$ or $Z$).
Events with a selected lepton which is non-prompt or fake are themselves labelled as fake and are treated as a contribution to the background.

\begin{figure}[tbp]
  \centering
  \includegraphics[width=.45\textwidth]{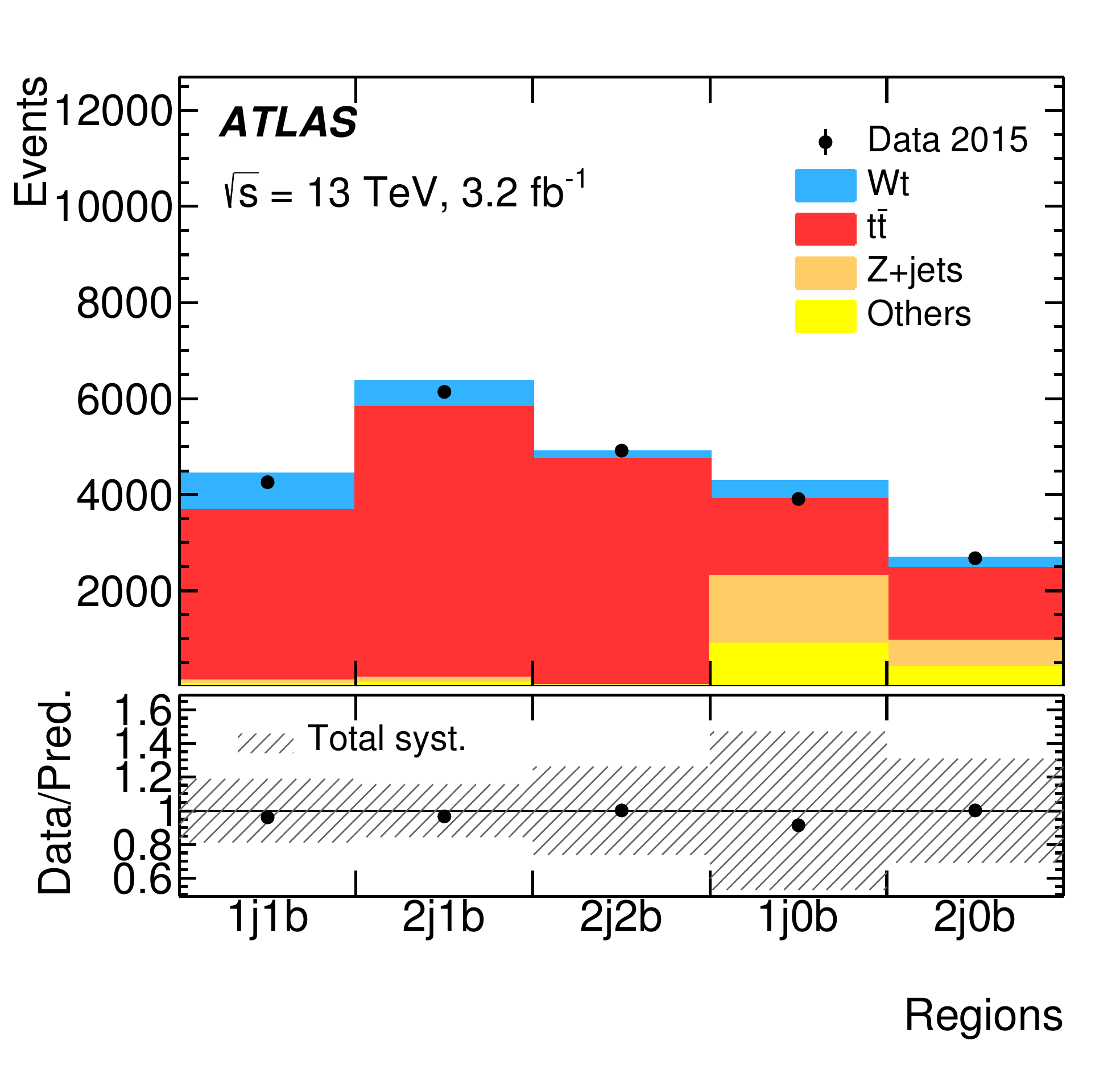}
  \caption{%
    Expected event yields for signal and backgrounds with their total systematic uncertainty (discussed in \Sect{\ref{sec:fit}}) and the number of observed events in the data are shown in the three fit regions (\texttt{1j1b}, \texttt{2j1b}, and \texttt{2j2b}) and the two additional regions (\texttt{1j0b} and \texttt{2j0b}).
The signal and backgrounds are normalised to their theoretical predictions, and the error bands represent the total systematic uncertainties which are used in this analysis.
  The upper panel gives the yields in number of events per bin, while the lower panel gives the ratios of the numbers of observed events to the total prediction in each bin.
  }

  \label{fig:regions}
\end{figure}

After this selection has been made, a further set of requirements is imposed with the aim of reducing the contribution from the \Zjets, diboson and fake/non-prompt lepton backgrounds.
The resultant sample is intended to consist almost entirely of $Wt$ signal and \ttbar{} background (a breakdown of the expected signal contributions and background compositions in all regions can be seen in \Fig{\ref{fig:regions}}), which are subsequently separated by the BDT analysis.
Events in which the two leptons have the same flavour and an invariant mass consistent with a $Z$ boson ($81<m_{\ell\ell}<\SI{101}{\GeV}$) are vetoed, as well as those with an invariant mass $m_{\ell\ell}<\SI{40}{\GeV}$.
Further requirements on \met\ and $m_{\ell\ell}$ are chosen based on the flavour of the selected leptons (as shown in \Tab{\ref{tab:selection}}).
Events with different-flavour leptons are required to have $\met > \SI{20}{\GeV}$, with the requirement raised to $\met>\SI{50}{\GeV}$ when the dilepton invariant mass satisfies $\mll<\SI{80}{\GeV}$.
All events with same-flavour leptons must satisfy $\met>\SI{40}{\GeV}$.
For same-flavour events, the \Zjets{} background is concentrated in a region of the \mll--\met plane corresponding to values of \mll near the $Z$ mass, and towards low values of \met.
Therefore, a selection in \met and \mll is used to remove these backgrounds: events with $\SI{40}{\GeV} < \mll < \SI{81}{\GeV}$ are required to satisfy $4 \times \met > 5 \times \mll$ while events with $\mll > \SI{101}{\GeV}$ are required to satisfy $2 \times \mll + \met > \SI{300}{\GeV}$.
The requirements for the same- and different-flavour events are chosen separately due to the kinematically different processes contributing to the \Zjets{} background, namely $Z\to ee/\mu\mu$ in same-flavour events and $Z\to\tau\tau$ in different-flavour events.
These requirements reduce the \Zjets contributions in the signal regions to \SI{12}{\%} according to simulation.
The partitioning of events into different selections based on lepton flavour, \met, and $m_{\ell\ell}$ is described well by the simulation, motivating the choice to merge these selection regions into the signal regions described below.

The sample of selected events is divided into regions based on the number of jets and $b$-tagged jets.
At LO, the signal process results in a final state with one $b$-jet arising from the top quark decay, and no additional jets, while the \ttbar{} process results in two $b$-jets from the top quark decays.
Events with additional jets are also studied since the underlying event, higher order QCD and other effects may produce additional jets in signal events.

\begin{figure}[tbp]
  \centering
  \includegraphics[width=.50\textwidth]{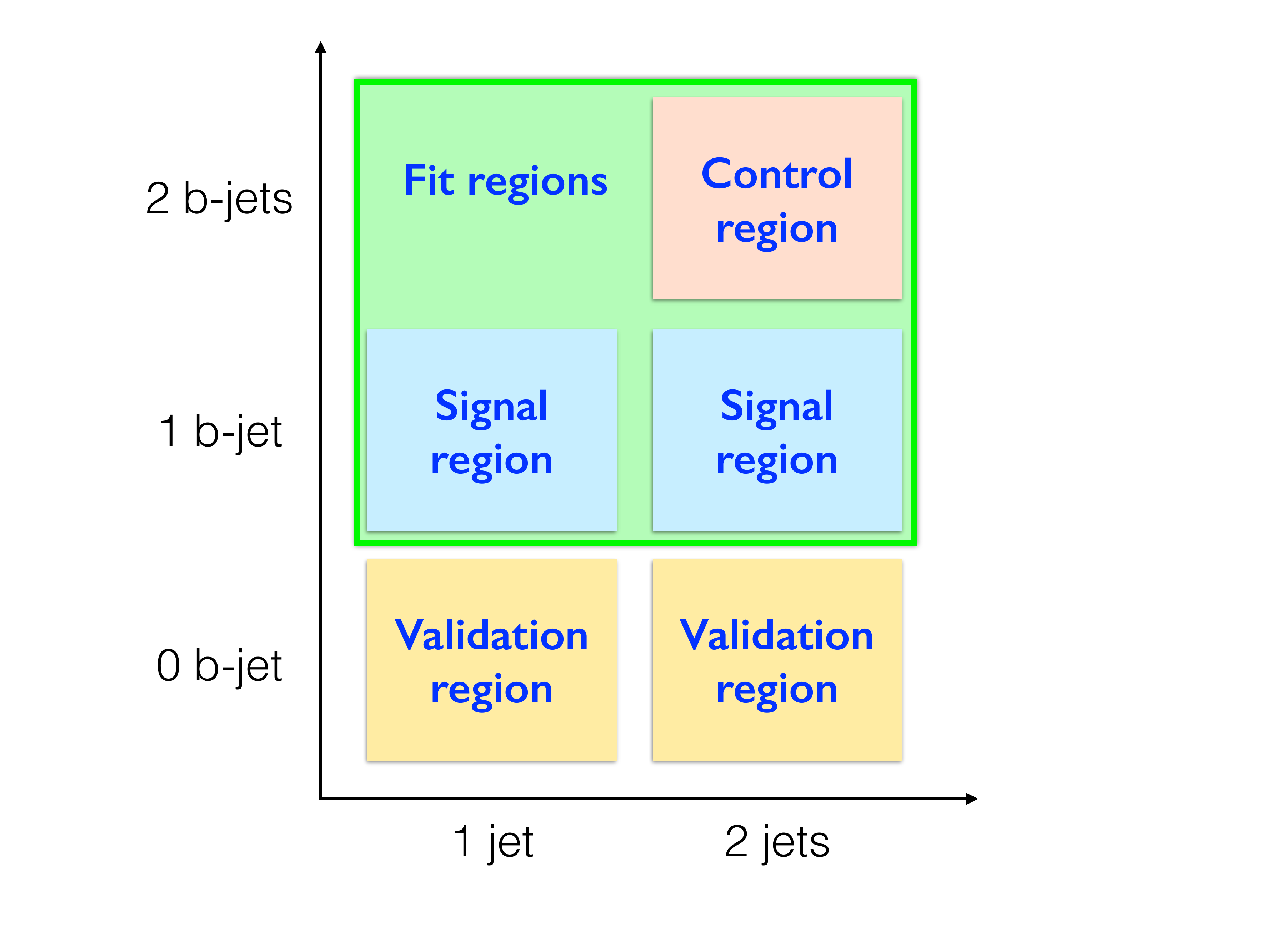}
  \caption{%
    A schematic view of signal, control and validation regions. Signal and control regions are used in the fit.
  }
  \label{fig:regions_schema}
\end{figure}

Corresponding to these expected final states, two signal regions are defined by the presence of exactly one $b$-tagged jet and either zero (denoted \texttt{1j1b}) or one (denoted \texttt{2j1b}) additional jet.
A \ttbar{}-enriched control region is defined by the presence of exactly two jets, which are both $b$-tagged (denoted \texttt{2j2b}).
This control region is used to constrain the \ttbar{} background normalisation, and is expected to contain only a small ($< \SI{1}{\%}$) proportion of signal events.
These three regions --- \texttt{1j1b}, \texttt{2j1b} and \texttt{2j2b} --- are called the fit regions, as they are used in the simultaneous fit described in \Sect{\ref{sec:fit}}.
The total efficiency in simulation to accept a dilepton $Wt$ signal event into one of the signal or control regions is about \SI{12}{\%}, while the efficiency to accept a dilepton \ttbar\ background event to the same regions is about \SI{5}{\%} estimated in simulation.
Event yields for each fit region are presented in \Sect{\ref{sec:results}}.
Two additional regions, in which events are required to contain one (denoted \texttt{1j0b}) or two (denoted \texttt{2j0b}) jets but no $b$-tagged jets are used to validate the description of the data by the simulation.
A schematic view of the regions definition is shown in \Fig{\ref{fig:regions_schema}}.

\begin{table}
  \renewcommand{\arraystretch}{1.3}
  \caption{Summary of event selection criteria used in the analysis.}
  \centering
  \begin{tabular}{crl}
    \toprule
    \multicolumn{3}{c}{At least one jet with $\pt>\SI{25}{\GeV}$, $|\eta|<2.5$} \\
    \multicolumn{3}{c}{Exactly two leptons of opposite charge with $\pt>\SI{20}{\GeV}$,} \\
    \multicolumn{3}{c}{$|\eta|<2.5$ for muons and $|\eta|<2.47$ excluding $1.37<|\eta|<1.52$ for electrons} \\
    \multicolumn{3}{c}{At least one lepton with $\pt>\SI{25}{\GeV}$, veto if third lepton with $\pt>\SI{20}{\GeV}$} \\
    \multicolumn{3}{c}{At least one lepton matched to the trigger object} \\
    \midrule
    \multirow{2}{*}{Different flavour}
      & $\met > \SI{50}{\GeV}$, & \hspace{1em}if $\mll < \SI{80}{\GeV}$ \\
      & $\met > \SI{20}{\GeV}$, & \hspace{1em}if $\mll > \SI{80}{\GeV}$ \\
    \midrule
    \multirow{5}{*}{Same flavour}
      & $\met > \SI{40}{\GeV}$, & \hspace{1em}always \\
      & veto, & \hspace{1em}if $\mll < \SI{40}{\GeV}$ \\
      & $4 \met > 5 \mll$, & \hspace{1em}if $\SI{40}{\GeV} < \mll < \SI{81}{\GeV}$ \\
      & veto, & \hspace{1em}if $\SI{81}{\GeV} < \mll < \SI{101}{\GeV}$ \\
      & $2 \mll + \met > \SI{300}{\GeV}$, & \hspace{1em}if $\mll > \SI{101}{\GeV}$ \\
    \bottomrule
  \end{tabular}
  \label{tab:selection}
\end{table}

\section{Separation of signal from background}
\label{sec:mva}
After the event selection is performed, the data sample consists primarily of \ttbar{} events with a significant number of $Wt$ signal events (see for example \Fig{\ref{fig:regions}}).
As there is no single observable that clearly discriminates between the $Wt$ signal and the \ttbar\ background, several observables are combined into a single discriminator using a BDT technique~\cite{bdt}.
A collection of decision trees is created that weakly separates events into signal and background based on a number of binary decisions considering a single observable at a time.
A boosting algorithm is then used to assign weights to each tree such that the ensemble of weak classifiers performs as a strong classifier~\cite{boosting}.
In this analysis, the BDT implementation is provided by the \textsc{tmva} package~\cite{tmva}, using the GradientBoost algorithm.

Separate BDTs are prepared for the analysis regions \texttt{1j1b} and \texttt{2j1b}.
Due to the low efficiency to accept a $Wt$ event in the \texttt{2j2b} region, the computing cost to simulate events in this region is especially large.
Since the expected gain in signal precision from subdividing the \texttt{2j2b} region is minimal, no BDT is constructed here and a single bin is used.
The BDTs are optimised to distinguish between $Wt$ and \ttbar{} by using the nominal $Wt$ MC sample, the alternative $Wt$ MC sample with diagram subtraction scheme and the nominal \ttbar{} MC sample; for each sample, half of the events are used for training while the other half is reserved for testing.
For each region, a large list of variables is prepared for the BDT.
An optimisation procedure is then carried out in each region to select a subset of input variables and a set of BDT parameters (such as the number of trees in the ensemble and the maximum depth of the individual decision trees).
The optimisation is designed to provide the best separation between the $Wt$ signal and \ttbar{} background while avoiding sensitivity to statistical fluctuations in the training sample.

The variables considered are derived from the kinematic properties of subsets of the selected physics objects defined in \Sect{\ref{sec:objects}} for each event.
For a set of objects $o_1$\ldots$o_n$:
\pTsys{}($o_1$\ldots$o_n$) is the transverse momenta of various subsets;
\HT($o_1$\ldots$o_n$) is the scalar sum of transverse momenta;
$\sum{\ET}$ is the scalar sum of the transverse momenta of all objects which contribute to the \met{} calculation;
$\sigma(\pTsys{})$ is the ratio of \pTsys to $(\HT + \sum{\ET})$;
$m$($o_1$\ldots$o_n$) is the invariant mass of various subsets;
$m_{\mathrm{T}}$($o_1$\ldots$o_n$) is the transverse mass (i.e. the sum of the invariant masses of $o_{1}$\ldots$o_{n}$ each projected onto the transverse plane); and
$E/m$($o_1$\ldots$o_n$) is the ratio of energy to invariant mass.
Two-dimensional vectors such as $\vec{E}_{\text{T}}^{\text{miss}}$ are assigned four-momenta by assuming zero mass and $z$-component.
For two systems of objects $s_{1}$ and $s_{2}$:
$\Delta R (s_1, s_2)$ is the separation in $\phi$--$\eta$ space;
$\Delta \pT (s_1, s_2)$ is the \pT difference;
$\Delta \phi (s_1, s_2)$ is the $\phi$ difference; and
$C(s_1, s_2)$, the centrality, is the ratio of the scalar sum of \pT to the sum of energy.

\begin{figure}[tp]
  \centering
  \begin{tabular}{cc}
    \includegraphics[width=.42\textwidth]{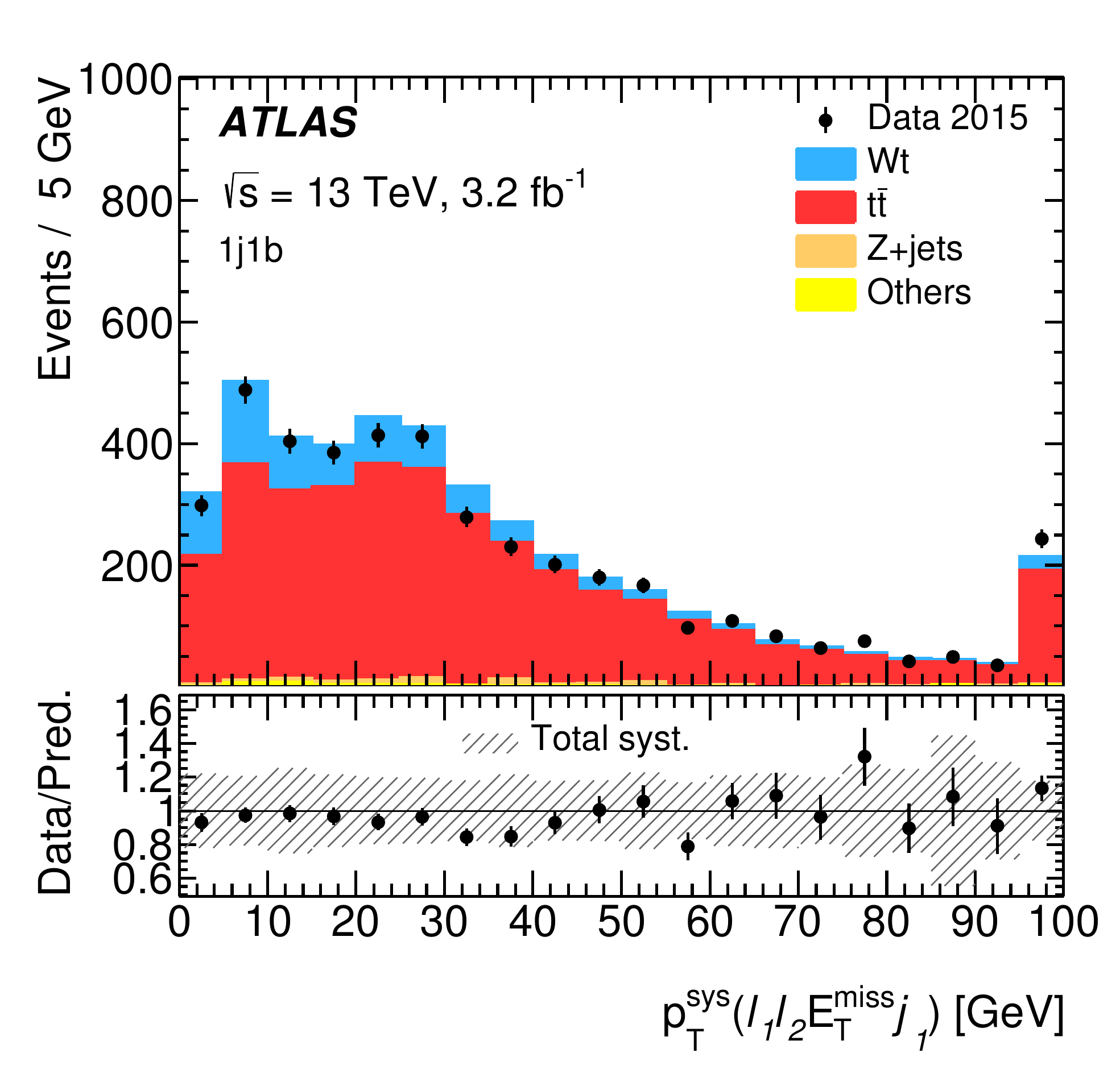} &
    \includegraphics[width=.42\textwidth]{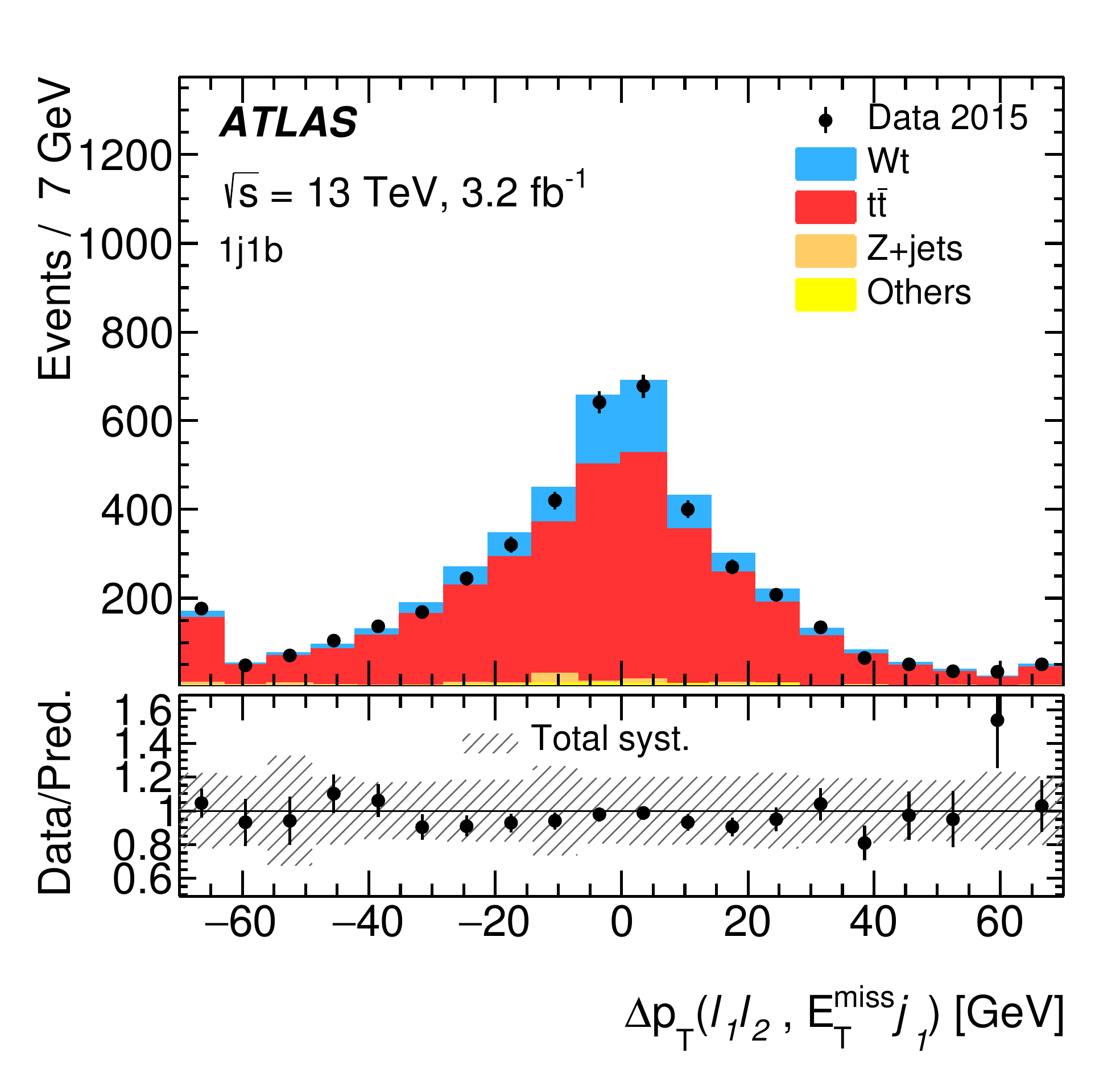} \\[-6pt]
    (a) & (b) \\

    \includegraphics[width=.42\textwidth]{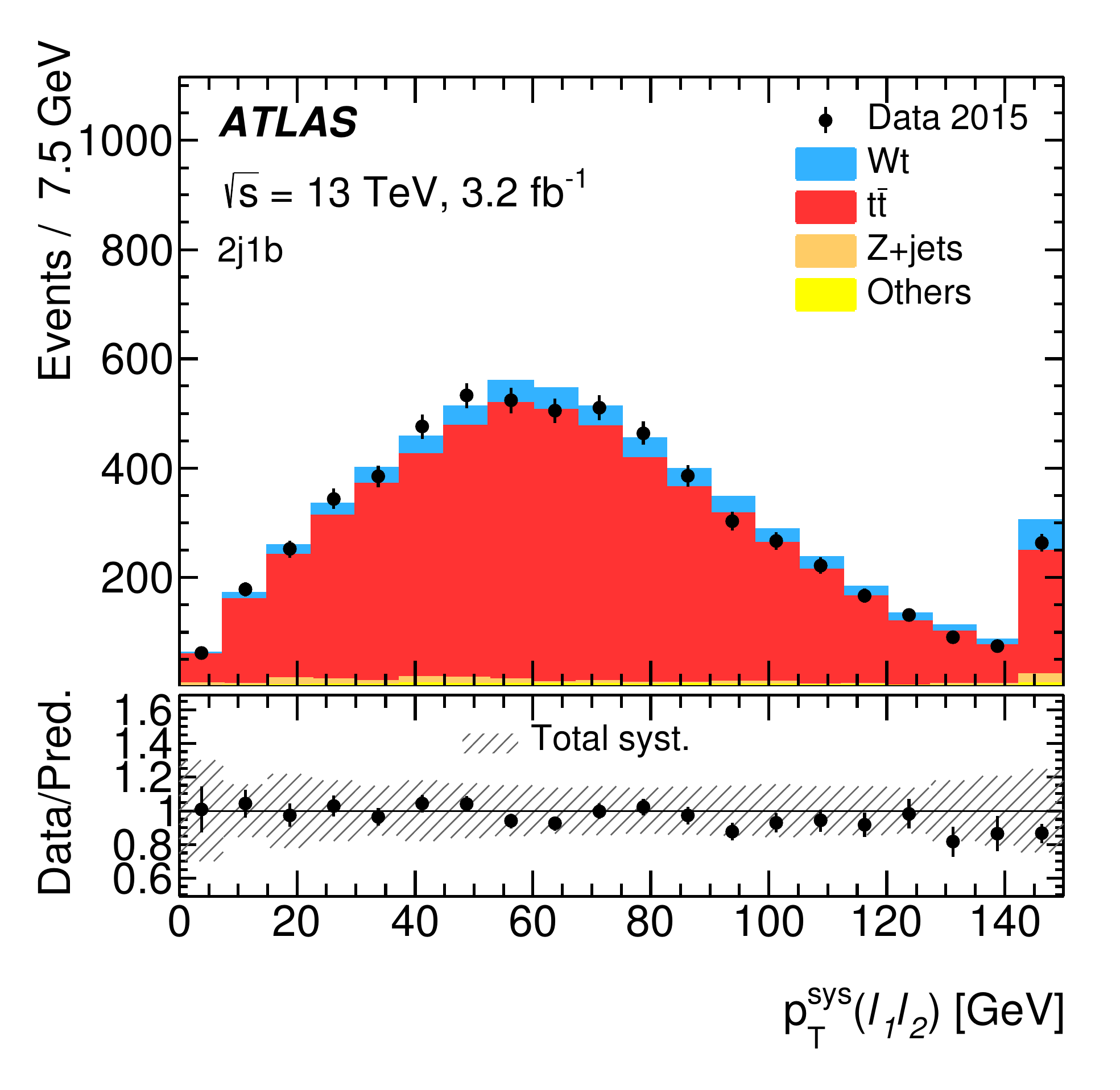} &
    \includegraphics[width=.42\textwidth]{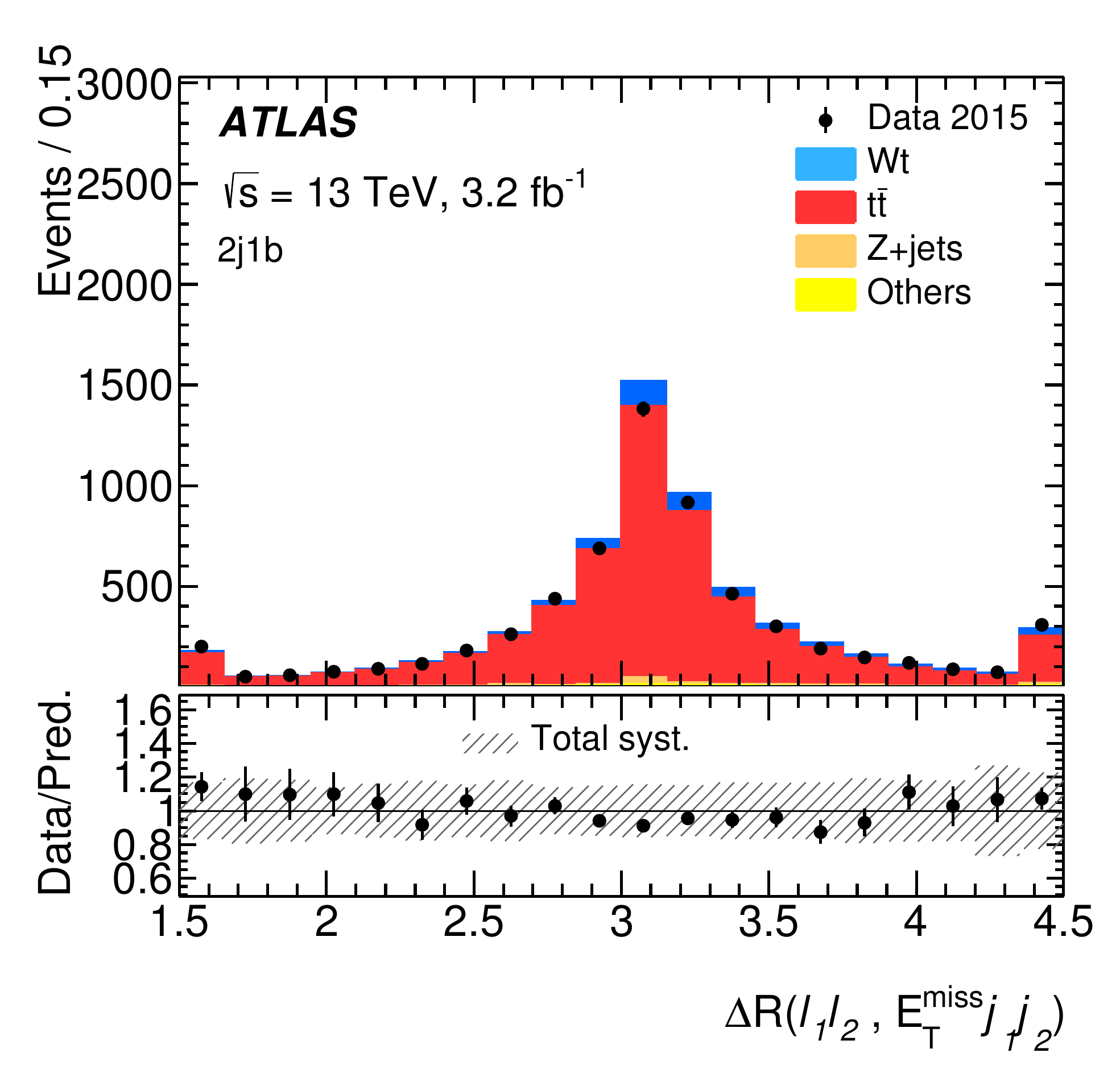} \\[-6pt]
    (c) & (d) \\

  \end{tabular}
  \caption{%
    Distributions of the two most powerful BDT input variables in each fit region:
    in the \texttt{1j1b} region 
    (a)  $\pTsys(\ell_{1} \ell_{2} \MET j_{1})$ and
    (b) $\Delta\pT(\ell_{1} \ell_{2} \MET j_{1})$;
    in the \texttt{2j1b} region
    (c) $\pTsys(\ell_{1}, \ell_{2})$ and
    (d) $\Delta R(\ell_{1} \ell_{2}, \MET j_{1} j_{2})$.
  The signal and backgrounds are normalised to their theoretical predictions, and the error bands represent the total systematic uncertainties in the Monte Carlo predictions.
  The first and last bins of each distribution contain overflow events.
  The upper panels give the yields in number of events per bin, while the lower panels give the ratios of the numbers of observed events to the total prediction in each bin.
}
  \label{fig:mva_variables}
\end{figure}

The final set of input variables used in each BDT is listed in \Tab{\ref{tab:mva_separation}} along with the separating power of each variable.%
\footnote{The separating power, $S$, is a measure of the difference between probability distributions of signal and background in the variable, and is defined as
  \begin{equation*}
    \langle S^2 \rangle = \frac{1}{2}  \int{\frac{(Y_{\text{s}}(y) - Y_{\text{b}}(y))^2}{(Y_{\text{s}}(y)+Y_{\text{b}}(y))} \text{d}y}
  \end{equation*}
where $Y_{\text{s}}(y)$ and $Y_{\text{b}}(y)$ are the signal and background probability distribution functions of each variable $y$, respectively.}
In order to check that the variables and their correlations in $Wt$ signal and the background events are well modelled by simulation, the distributions of these variables and the BDTs are compared between the MC prediction and the observed data, using a Kolmogorov--Smirnov (KS) statistical test~\cite{kstest} to check agreement.
The distributions of the two most powerful variables in each fit region are shown in \Fig{\ref{fig:mva_variables}}.
The MC predictions describe the data well, within the total systematic uncertainties.

\begin{table}[thbp!]
  \centering
  \caption{%
    The variables used in each BDT and their separating powers (a measure of the difference between probability distributions of signal and background in the variable, denoted $S$).
    The variables are derived from the four-momenta of the leading (sub-leading) lepton $\ell_1$ ($\ell_2$), the leading (sub-leading) jet $j_1$ ($j_2$) and \met{}.
    The last row gives the separation power of the BDT discriminant output.
  }
  {\footnotesize
  \begin{minipage}[t][][b]{0.32\textwidth}
    \centering
    \renewcommand{\arraystretch}{1.3}
    \begin{tabular}{lr}
	\toprule
	\multicolumn{2}{c}{\texttt{1j1b}} \\
	Variable & \multicolumn{1}{c}{$S$ $\left[ 10^{-2} \right]$} \\
	\midrule

$\pTsys       (\ell_{1} \ell_{2} \MET j_{1})   $ & 5.3 \\
$\Delta\pT    (\ell_{1} \ell_{2}, \MET j_{1})  $ & 2.9 \\
$\sum{\ET}                                     $ & 2.7 \\
$\Delta\pT    (\ell_{1} \ell_{2}, \MET)        $ & 1.2 \\
$\pTsys       (\ell_{1} \MET j_{1})            $ & 0.9 \\
$C            (\ell_{1} \ell_{2})              $ & 0.9 \\
$\Delta\pT    (\ell_{1}, \MET)                 $ & 0.8 \\
  \midrule
  BDT discriminant                                     & 8.6 \\

        \bottomrule
\end{tabular}

  \end{minipage}
  \begin{minipage}[t][][b]{0.32\textwidth}
    \centering
    \renewcommand{\arraystretch}{1.3}
    \begin{tabular}{lr}
	\toprule
	\multicolumn{2}{c}{\texttt{2j1b}} \\
	Variable & \multicolumn{1}{c}{$S$ $\left[ 10^{-2} \right]$} \\
	\midrule

$\pTsys          (\ell_{1} \ell_{2})                       $ & 1.7 \\
$\Delta R	 (\ell_{1} \ell_{2}, \MET j_{1} j_{2}) 	   $ & 1.7 \\
$\Delta R        (\ell_{1} \ell_{2}, j_{1} j_{2})          $ & 1.5 \\
$m	 	 (\ell_{1} j_{2}) 	                   $ & 1.4 \\
$\Delta\pT       (\ell_{1} \ell_{2}, \MET)                 $ & 1.4 \\
$\Delta\pT       (\ell_{1}, j_{1})                         $ & 1.4 \\
$m               (\ell_{1} j_{1})                          $ & 1.3 \\
$\pT             (\ell_{1})                                $ & 1.3 \\
$\sigma(\pTsys)	 (\ell_{1} \ell_{2} \MET j_{1}) 	   $ & 1.2 \\
$\Delta R	 (\ell_{1}, j_{1}) 	                   $ & 1.2 \\
$\pT             (j_{2})                                   $ & 0.9 \\
$\sigma(\pTsys)	 (\ell_{1} \ell_{2} \MET j_{1} j_{2}) 	   $ & 0.9 \\
$m               (\ell_{2} j_{1} j_{2})                    $ & 0.3 \\
$m	 	 (\ell_{2} j_{1}) 	                   $ & 0.3 \\
$m	 	 (\ell_{2} j_{2}) 	                   $ & 0.1 \\
  \midrule
  BDT discriminant                                     & 10.9 \\

	\bottomrule
\end{tabular}

  \end{minipage}
  }
  \label{tab:mva_separation}
\end{table}

\section{Systematic uncertainties}
\label{sec:systematics}
Systematic uncertainties are divided into experimental and theoretical sources.
Each uncertainty is assigned a Gaussian-constrained nuisance parameter, which allows the uncertainty to be constrained by data.

The experimental sources of uncertainty include the measurement of the luminosity, lepton efficiency scale factors used to correct simulation to data, lepton energy scale and resolution, \MET soft-term calculation, jet energy scale and resolution, and the \btag efficiency.
Among these, the dominant sources of uncertainty are due to the determination of the jet energy scale (JES) and jet energy resolution.
\Tab{\ref{tab:systematics}} gives a breakdown of uncertainties in the final fitted cross-section.

The JES uncertainty~\cite{ATL-PHYS-PUB-2015-015} is divided into a total of 18 components, which are derived using $\sqrt{s}=\SI{13}{\TeV}$ data.
The uncertainties from \textit{in situ} analyses including studies of $Z/\gamma +$jet and dijet events are represented with six orthogonal components (JES Eff1--6).
The full description of jet uncertainties and correlations is reduced to obtain this set of uncertainty components that can be used as nuisance parameters in a likelihood fit.
This is done by diagnoalising the covariance matrix describing the jet uncertainties to obtain a set of reduced uncertainties corresponding to the eigenvector--eigenvalue pairs as demonstrated in Ref.~\cite{jesdecomposition}.
Other components are model uncertainties (such as flavour composition, $\eta$ intercalibration model), and other systematics in the JES determination (such as pile-up jet area $\rho$).
The most significant JES uncertainty components for this analysis are the \textit{in situ} calibration and the flavour composition uncertainty, which is the dependence of the jet calibration on the fraction of quark or gluon jets in data.
The jet energy resolution uncertainty estimate~\cite{ATL-PHYS-PUB-2015-015} is based on comparisons of simulation and data using \textit{in situ} studies with Run-1 data.
These studies are then cross-calibrated and checked to confirm good agreement with Run-2 data.

As discussed in \Sect{\ref{sec:objects}}, the \MET calculation includes contributions from hard sources, including leptons and jets, in addition to soft terms which arise primarily from low-\pt pile-up jets and underlying-event activity.
The uncertainty associated with the hard terms is propagated from the corresponding uncertainties in the energy/momentum scales and resolutions for jets and leptons, and is classified together with the uncertainty associated with the hard objects.
The uncertainty associated with the soft term is estimated by comparing the simulated scale and resolution to that in data, including differences in uncertainties due to model dependence.

Uncertainties in the scale factors to correct the \btag efficiency in simulation to the efficiency in data are assessed using independent eigenvectors for the efficiency of $b$-jets, $c$-jets, light-parton jets, and two extrapolation uncertainty factors.
These \btag uncertainties are determined with $\sqs=\SI{13}{\TeV}$ data for $b$-jets, while for $c$-jets and light-parton jets they are determined in $\sqs=\SI{8}{\TeV}$ data, then extrapolated to and checked with $\sqs=\SI{13}{\TeV}$ data.
The extrapolation is performed by adding additional uncertainties to cover changes made in the inner detector and tracking algorithms between Run-1 and Run-2 data, accounting for fake tracks, tracking efficiency and tracking resolution.
These $c$-jet and light-parton jet scale factors were later checked against similar scale factors derived on $\sqs=\SI{13}{\TeV}$ data, and the scale factors with added uncertainties were found to agree well with the full run-2 scale factors.

Systematic uncertainties in lepton momentum resolution and scale, trigger efficiency, isolation efficiency, and identification efficiency are also considered.
These uncertainties arise from corrections to simulation based on studies of $Z \to ee$ and $Z \to \mu\mu$ data.
In this analysis the effects of the uncertainties in these corrections are relatively small.

A \SI{2.1}{\%} uncertainty is assigned to the integrated luminosity determination for 2015 data.
It is derived, following a methodology similar to that detailed in \Ref{\cite{DAPR-2013-01}}, from a calibration of the luminosity scale using $x$--$y$ beam-separation scans performed in August 2015.

Uncertainties stemming from theoretical models are estimated by comparing a set of predicted distributions produced with different assumptions and applying the difference observed as a weight to the nominal $Wt$ or \ttbar distribution.
The main uncertainties are due to the NLO matrix element (ME) generator, parton shower and hadronisation generator, initial- and final-state radiation (I/FSR) tuning and the PDF.
The NLO matrix element uncertainty is estimated by comparing two NLO matching methods: the predictions of \POWHEGBOX and \MGMCatNLO, both interfaced with \HERWIGpp.
The parton shower, hadronisation, and underlying-event model uncertainty is estimated by comparing \POWHEGBOX interfaced with either \PYTHIA6 or \textsc{Herwig++}.
The uncertainty from the matrix element generator is treated as uncorrelated between the $Wt$ and \ttbar processes, while the uncertainty from the parton shower generator is treated as correlated.
The I/FSR tuning uncertainty is estimated by taking half of the difference between samples with \POWHEGBOX interfaced with \PYTHIAV{6} tuned with either more or less radiation, and is uncorrelated between the $Wt$ and \ttbar processes.
The choice of scheme to account for the interference between the $Wt$ and \ttbar{} processes constitutes another source of systematic uncertainty for the signal modelling, and it is estimated by comparing samples using either the diagram removal scheme or the diagram subtraction scheme, both generated with \POWHEGBOX{}+\PYTHIAV{6}.
The uncertainty due to the choice of PDF is estimated using the PDF4LHC15 combined PDF set~\cite{Butterworth:2015oua}.
The difference between the central CT10~\cite{ct10} prediction and the central PDF4LHC15 prediction (PDF central value) is taken and symmetrised together with the internal uncertainty set provided with PDF4LHC15.
For \ttbar{} and $Wt$ modelling, the NLO matrix element model, parton shower model, and PDF uncertainties are estimated using fast-simulated samples; for $Wt$, fast simulation is also used for I/FSR.
In each case where results from two samples must be compared, fast simulated samples are only compared to other fast simulated samples.

Additionally, normalisation uncertainties of \SI{100}{\%} are assumed for the fake/non-prompt lepton backgrounds.
The \Zjets backgrounds with one $b$-tagged jet are assigned a \SI{50}{\%} uncertainty, while a \SI{100}{\%} uncertainty is assumed for \Zjets events with two $b$-tagged jets.
These uncertainties are chosen to be consistent with previous ATLAS studies of these processes in dedicated validation regions.
Diboson backgrounds are assigned an uncertainty of \SI{25}{\%} to cover the difference between the predictions of the \SHERPA and \POWHEGBOX generators.
These uncertainties are treated as uncorrelated across the various regions of jet and $b$-tagged jet multiplicity.

\section{Extraction of signal cross-section}
\label{sec:fit}
The $Wt$ cross-section is extracted from the data using a profile-likelihood fit that combines inputs from each signal and control region to constrain backgrounds and systematic uncertainties.
The fit uses the \texttt{HistFitter}~\cite{histfitter} software framework, which is in turn built on the \texttt{HistFactory}, \texttt{RooStats}, and \texttt{RooFit}~\cite{histfactory} frameworks.

The fit uses the binned BDT response for MC events in two of the three fit regions (\texttt{1j1b} and \texttt{2j1b}) and a single bin in the \texttt{2j2b} region to construct templates for the $Wt$ signal and each modelled background (\ttbar, \Zjets, diboson, fake or non-prompt leptons).
For each signal and background template, an additional template is constructed for each of the MC sample variations (see \Sect{\ref{sec:systematics}}) accounting for a systematic uncertainty.
Systematic uncertainties are considered by allowing Gaussian-constrained nuisance parameters to deform fit templates while simultaneously varying the normalisation of the templates.
The normalisation of the \ttbar{} background, $\mu_{\ttbar}$, is also determined in the fit by assigning an unconstrained parameter to the \ttbar{} normalisation.
Other backgrounds are constrained within their systematic uncertainties by Gaussian-constrained nuisance parameters, and all templates are affected by the overall luminosity uncertainty.

A global likelihood function is constructed to describe the level of agreement between data and prediction as a function of the parameter of interest, namely the $Wt$ signal strength $\mu_{Wt}$, and a list of nuisance parameters each describing the influence of a different source of systematic uncertainty.
The $Wt$ cross-section and its uncertainty are extracted from the fitted value of $\mu_{Wt}$, with a value of unity corresponding to the predicted NLO+NNLL $\sigma_{\text{theory}}$ value.

\begin{table}[htbp]
  \centering
  \caption{Relative uncertainties in the $Wt$ cross-section.
    These are estimated by fixing each uncertainty parameter to its post-fit $\pm 1 \sigma$ uncertainties, re-fitting, and assessing the change in the signal strength.
    Due to correlations between parameters, the individual uncertainty categories are not expected to add up to the total systematic uncertainty.
    The statistical uncertainty is evaluated by fitting without any nuisance parameters corresponding to systematic uncertainties in the fit, and the total systematic uncertainty
    is evaluated by subtracting the statistical uncertainty from the total uncertainty in quadrature.
  }

  \sisetup{}
  \begin{tabular}{l
    S[round-mode=figures, round-precision=2]}
    \toprule
    Source   & $\Delta \sigma_{Wt} / \sigma_{Wt} [\%]$ \\
    \midrule
    Jet energy scale             & 21.0 \\
    Jet energy resolution        & 8.6 \\
    \met soft terms              & 5.3 \\
    $b$-tagging                  & 4.3 \\
    Luminosity                   & 2.3 \\
    Lepton efficiency, energy scale and resolution  & 1.3 \\
    \midrule
    NLO matrix element generator       & 18.4  \\
    Parton shower and hadronisation & 7.1 \\
    Initial-/final-state radiation  & 6.4 \\
    Diagram removal/subtraction  & 5.3 \\
    Parton distribution function   & 2.7 \\
    Non-\ttbar background normalisation     & 3.7 \\
    \midrule
    Total systematic uncertainty & 29.6 \\
    Data statistics              & 10.2 \\
    \midrule
    Total uncertainty            & 31.3 \\
    \bottomrule
  \end{tabular}

  \label{tab:systematics}
\end{table}

\section{Results}
\label{sec:results}
The expected and fitted yields from data are measured in the three fit regions.
The majority of signal events fall in the \texttt{1j1b} and \texttt{2j1b} regions, with the former giving the better signal to background ratio as well as the larger yield of signal events.
\Tab{\ref{tab:fit_yields}} shows the fitted yields of each process.
From the fitted $Wt$ yield, a cross-section is then extracted.
The result is a measured cross-section of \fullXsSplit{}, corresponding to an observed (expected) significance of \signifObserved{} (\signifExpected{}).
Most pairs of parameters in the fit show small correlations, generally at the \SI{25}{\%} level or less.
The most correlated pairs of nuisance parameters are the modelling uncertainties due to matrix element and parton shower (\SI{57}{\%}), and the parameters related to JES flavour composition and parton shower uncertainties (\SI{45}{\%}).

\begin{figure}[htbp]
  \centering
  \includegraphics[width=0.6\textwidth]{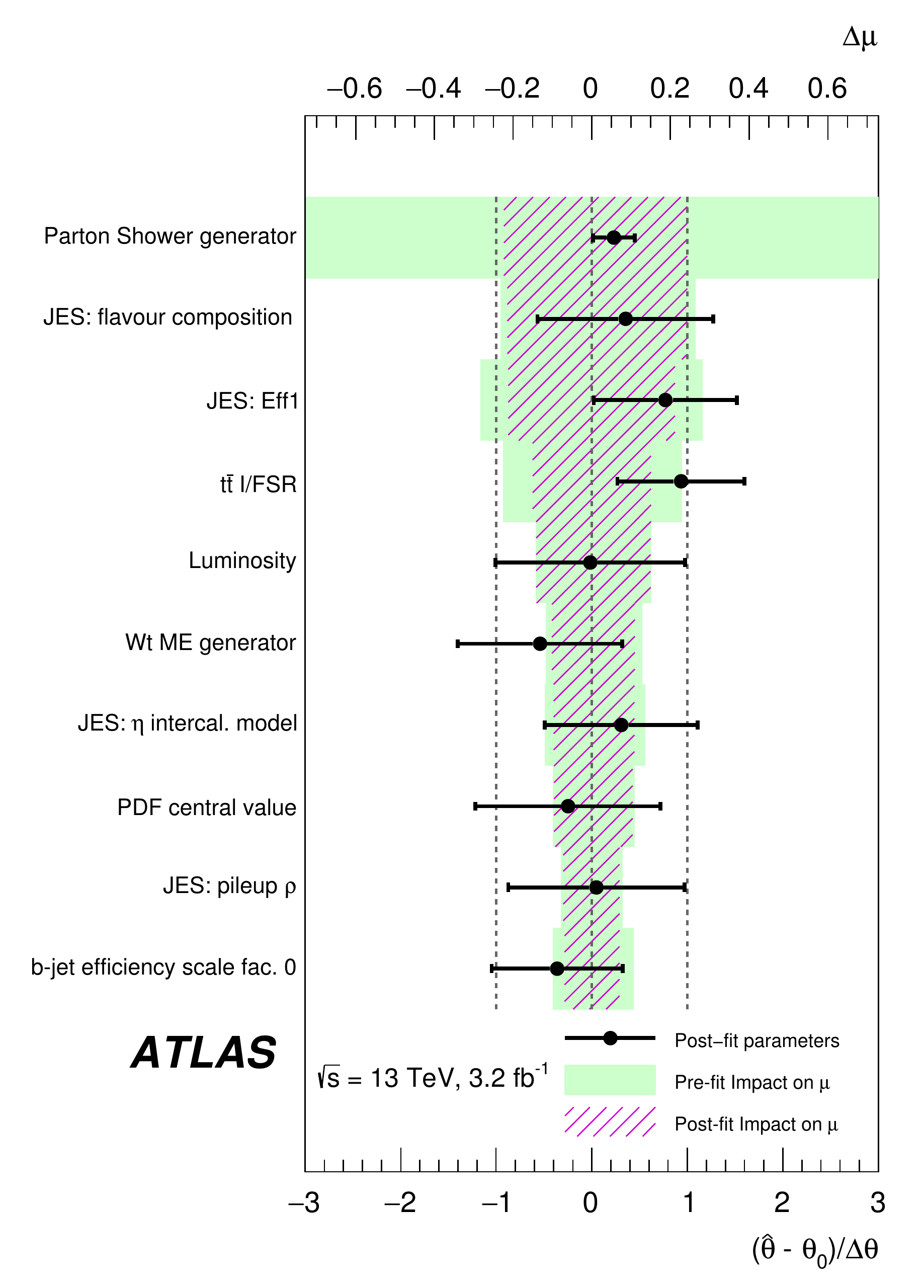}
  \caption{%
    List of fit parameters ranked by post-fit impact on the signal strength.
    The fit parameters ($\theta$) here correspond to the nuisance parameters from \Sect{\ref{sec:fit}}.
    Impact ($\Delta \mu$) is calculated by fixing the parameter to its $\pm1 \sigma$ values, fixing all other parameters to their nominal values, re-fitting the signal strength, and evaluating the change in signal strength with respect to the nominal fit.
    Green bands indicate the impacts computed with $\sigma$ corresponding to the pre-fit uncertainty, and hatched purple bands indicate the impacts computed with $\sigma$ corresponding to the post-fit uncertainty.
    The black points represent $(\hat{\theta} - \theta_0)/\Delta \theta$, the shifted and scaled post-fit parameter values, while the error bars are the post-fit errors of the fit parameter.
    The meanings of the labels and abbreviations are detailed in \Sect{\ref{sec:systematics}}.
}
  \label{fig:test_pulls}
\end{figure}

\Fig{\ref{fig:test_pulls}} shows the fit parameters ($\theta$) with the highest post-fit impact on the signal strength, and also gives the pre-fit impacts as well as post-fit parameter values.
The post-fit parameters, $\theta$, are shifted and re-scaled by $(\hat{\theta} - \theta_0)/\Delta \theta$, where $\theta_0$ and $\hat{\theta}$ are the pre- and post-fit values of $\theta$, while $\Delta \theta$ is the pre-fit uncertainty on $\theta$.
Here the impact ($\Delta \mu$) of a parameter is defined as the change in signal strength observed when fixing this parameter to its $\pm1 \sigma$ values, fixing all other parameters to their nominal values, and fitting the signal strength.
The change is taken with respect to the nominal pre-fit value for pre-fit impact and with respect to the nominal post-fit value for the post-fit impact.
The pre-fit and post-fit impact are differentiated based on whether pre-fit or post-fit values of $\pm1 \sigma$ variations are assumed for the parameter under consideration.
The parameters with the highest post-fit impact are jet energy scale uncertainties and the modelling uncertainties due to parton shower and \ttbar initial- and final-state radiation.
Some parameters fit to values which are significantly different from unity; \ttbar\ initial- and final-state radiation and the component JES Eff1 each exhibit this effect.
This behaviour is expected as a few parameters could be pulled outside of the $\pm1 \sigma$ band when there are a large number of parameters being fitted,
while the majority should fall within the $\pm1 \sigma$ range.
Certain parameters are assigned post-fit uncertainties significantly smaller than the nominal pre-fit uncertainty values and are thus profiled or constrained by the observed data.
For example, the uncertainty due to parton shower generator would be among the most dominant uncertainties without the constraints from profiling, with pre-fit impacts exceeding \SI{60}{\%} of the signal strength.
However, information from the \texttt{2j2b} region about the \ttbar\ normalisation and the relative yields in the signal regions significantly constrain these uncertainties.
Another feature observed in this plot is how the sum in quadrature of the individual impacts is substantially smaller than the final uncertainty shown in \Tab{\ref{tab:systematics}}.
This is due to the correlations between the uncertainties, and in particular to the constraint provided by the \ttbar normalisation.
Some of the larger uncertainties such as the parton shower generator uncertainty have an asymmetric impact on the signal strength.
These asymmetries have been traced to originate from the large normalisation uncertainty on the \ttbar background.

\begin{figure}[htbp]
  \centering
  \includegraphics[width=\textwidth]{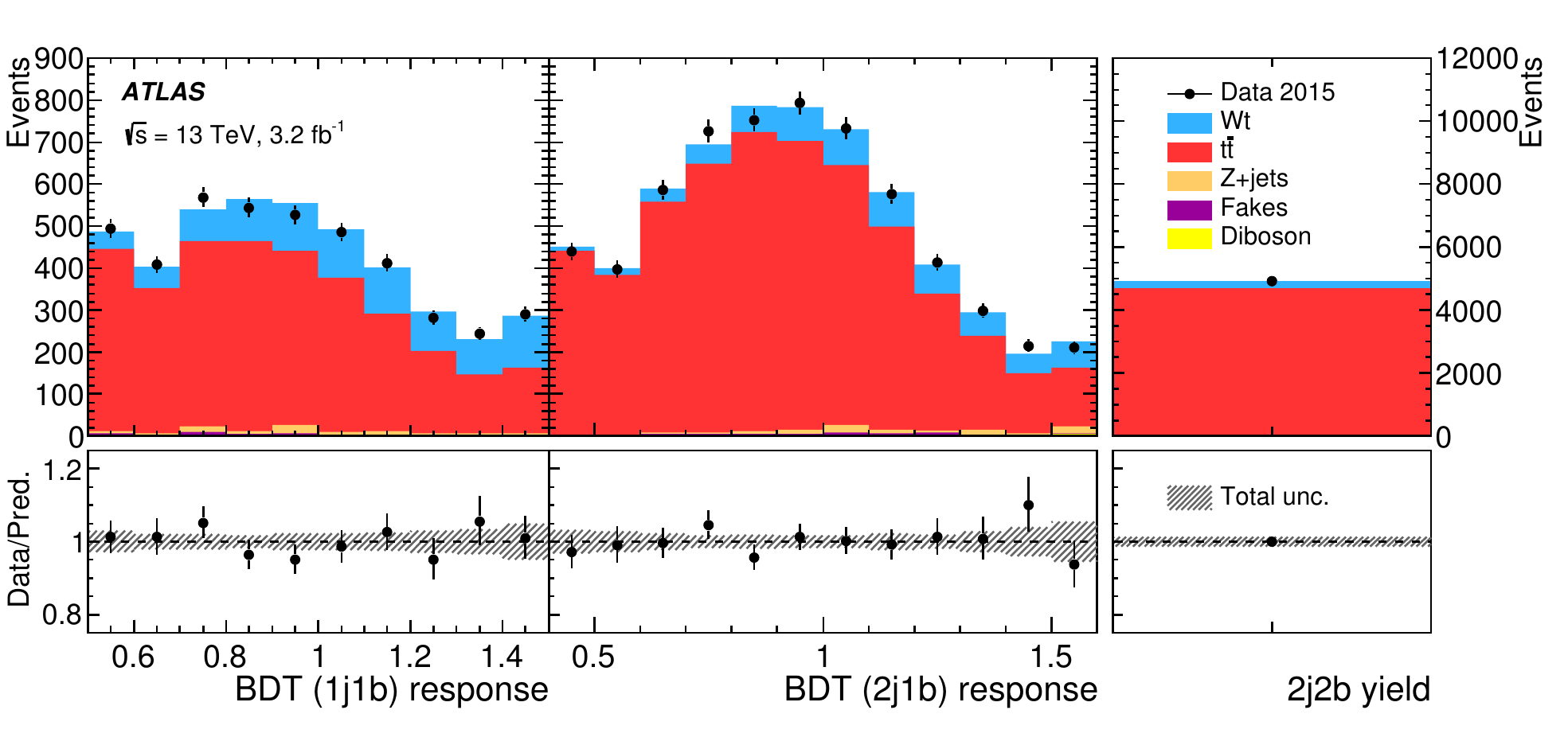}
  \caption{%
Post-fit distributions in the signal and control regions
\texttt{1j1b},
 \texttt{2j1b}, and
 \texttt{2j2b}.
The error bands represent the total uncertainties in the fitted results.
  The upper panels give the yields in number of events per bin, while the lower panels give the ratios of the numbers of observed events to the total prediction in each bin.
}
  \label{fig:afterfit_data}
\end{figure}

The MC predictions and data yields for the BDT response after setting all fit parameters to their final best-fit values are shown in \Fig{\ref{fig:afterfit_data}}, with error bands representing the total uncertainties in the fitted results.
The NLO+NNLL cross-section prediction agrees well with the measured value, and $\mu_{\ttbar}$ is fitted to $0.98\pm0.05$.

\begin{table}[htbp]
  \centering
  \caption{%
    Fit results for an integrated luminosity of \lumi.
    The errors shown are the final fitted uncertainties in the yields,
    including uncertainties in the fitted signal strength, systematic uncertainties, and
    statistical uncertainties, taking into account correlations and
    constraints induced by the fit.
  }
  \sisetup{group-separator = {}}
  {
  \begin{tabular}{l
    S[table-format=4.1]@{$\,\pm\,$}S[table-format=3.1]
    S[table-format=4.1]@{$\,\pm\,$}S[table-format=3.1]
    S[table-format=4.2]@{$\,\pm\,$}S[table-format=3.2]}
  	\toprule
  	                          & \multicolumn{2}{c}{\texttt{1j1b}} & \multicolumn{2}{c}{\texttt{2j1b}} & \multicolumn{2}{c}{\texttt{2j2b}} \\
  	\midrule
  	Observed events           & \multicolumn{1}{S[table-format=4.0]@{$\,\phantom{\pm}\,$}}{4254}    &
                              & \multicolumn{1}{S[table-format=4.0]@{$\,\phantom{\pm}\,$}}{6138}    &
                              & \multicolumn{1}{S[table-format=4.0]@{$\,\phantom{\pm}\,$}}{4912}    &  \\
  	\midrule
    Fitted events             & \multicolumn{1}{S[table-format=4.0]@{$\,\phantom{\pm}\,$}}{4257}    &
                              & \multicolumn{1}{S[table-format=4.0]@{$\,\phantom{\pm}\,$}}{6139}    &
                              & \multicolumn{1}{S[table-format=4.0]@{$\,\phantom{\pm}\,$}}{4908}    &  \\
  	\midrule

    Fitted $Wt$ events          &  910   & 210        & 640   & 160          &  210    &  82              \\
    Fitted \ttbar events        & 3230   & 210       & 5340   & 160          & 4670    & 110              \\
    Fitted \Zjets events        &   69   &  35         & 87   &  46          &    7.6  &   7.5            \\
    Fitted fake events          &   30   &  26         & 40   &  38          &   15    &  14              \\
    Fitted diboson events       &   23.5 &   6.0       & 24.8 &   6.2        &    0.91 &   0.23           \\
  	\bottomrule
  \end{tabular}
}
\label{tab:fit_yields}
\end{table}

\FloatBarrier
\section{Conclusion}
\label{sec:conclusion}

The inclusive cross-section for the associated production of a $W$ boson and top quark is measured using \lumi of $pp$ collision data collected at $\sqs=\SI{13}{\TeV}$ by the ATLAS detector at the LHC.
The analysis uses dilepton events with at least one $b$-tagged jet.
Events are separated into signal and control regions based on the number of jets and $b$-tagged jets, and the $Wt$ signal is separated from the \ttbar background using a BDT discriminant.
The cross-section is extracted by fitting templates to the BDT output distribution, and is measured to be \fullXsSplit{}. 
The measured value is in good agreement with the SM prediction of \ensuremath{\sigma_{\text{theory}}=\xspace{\SI[parse-numbers=false]{71.7\pm1.8\,(\mathrm{scale})\pm{3.4}\,(\mathrm{PDF})}{\fullXsUnit}}}~\cite{Kidonakis:2015}.

\section*{Acknowledgements}


We thank CERN for the very successful operation of the LHC, as well as the
support staff from our institutions without whom ATLAS could not be
operated efficiently.

We acknowledge the support of ANPCyT, Argentina; YerPhI, Armenia; ARC, Australia; BMWFW and FWF, Austria; ANAS, Azerbaijan; SSTC, Belarus; CNPq and FAPESP, Brazil; NSERC, NRC and CFI, Canada; CERN; CONICYT, Chile; CAS, MOST and NSFC, China; COLCIENCIAS, Colombia; MSMT CR, MPO CR and VSC CR, Czech Republic; DNRF and DNSRC, Denmark; IN2P3-CNRS, CEA-DSM/IRFU, France; SRNSF, Georgia; BMBF, HGF, and MPG, Germany; GSRT, Greece; RGC, Hong Kong SAR, China; ISF, I-CORE and Benoziyo Center, Israel; INFN, Italy; MEXT and JSPS, Japan; CNRST, Morocco; NWO, Netherlands; RCN, Norway; MNiSW and NCN, Poland; FCT, Portugal; MNE/IFA, Romania; MES of Russia and NRC KI, Russian Federation; JINR; MESTD, Serbia; MSSR, Slovakia; ARRS and MIZ\v{S}, Slovenia; DST/NRF, South Africa; MINECO, Spain; SRC and Wallenberg Foundation, Sweden; SERI, SNSF and Cantons of Bern and Geneva, Switzerland; MOST, Taiwan; TAEK, Turkey; STFC, United Kingdom; DOE and NSF, United States of America. In addition, individual groups and members have received support from BCKDF, the Canada Council, CANARIE, CRC, Compute Canada, FQRNT, and the Ontario Innovation Trust, Canada; EPLANET, ERC, ERDF, FP7, Horizon 2020 and Marie Sk{\l}odowska-Curie Actions, European Union; Investissements d'Avenir Labex and Idex, ANR, R{\'e}gion Auvergne and Fondation Partager le Savoir, France; DFG and AvH Foundation, Germany; Herakleitos, Thales and Aristeia programmes co-financed by EU-ESF and the Greek NSRF; BSF, GIF and Minerva, Israel; BRF, Norway; CERCA Programme Generalitat de Catalunya, Generalitat Valenciana, Spain; the Royal Society and Leverhulme Trust, United Kingdom.

The crucial computing support from all WLCG partners is acknowledged gratefully, in particular from CERN, the ATLAS Tier-1 facilities at TRIUMF (Canada), NDGF (Denmark, Norway, Sweden), CC-IN2P3 (France), KIT/GridKA (Germany), INFN-CNAF (Italy), NL-T1 (Netherlands), PIC (Spain), ASGC (Taiwan), RAL (UK) and BNL (USA), the Tier-2 facilities worldwide and large non-WLCG resource providers. Major contributors of computing resources are listed in Ref.~\cite{ATL-GEN-PUB-2016-002}.

\printbibliography

\newpage
\begin{flushleft}
{\Large The ATLAS Collaboration}

\bigskip

M.~Aaboud$^\textrm{\scriptsize 137d}$,
G.~Aad$^\textrm{\scriptsize 88}$,
B.~Abbott$^\textrm{\scriptsize 115}$,
J.~Abdallah$^\textrm{\scriptsize 8}$,
O.~Abdinov$^\textrm{\scriptsize 12}$,
B.~Abeloos$^\textrm{\scriptsize 119}$,
R.~Aben$^\textrm{\scriptsize 109}$,
O.S.~AbouZeid$^\textrm{\scriptsize 139}$,
N.L.~Abraham$^\textrm{\scriptsize 151}$,
H.~Abramowicz$^\textrm{\scriptsize 155}$
% ...
\bigskip
\\
% ...
$^{8}$ Department of Physics, The University of Texas at Arlington, Arlington TX, United States of America\\
% ...
$^{12}$ Institute of Physics, Azerbaijan Academy of Sciences, Baku, Azerbaijan\\
% ...
$^{88}$ CPPM, Aix-Marseille Universit{\'e} and CNRS/IN2P3, Marseille, France\\
% ...
$^{109}$ Nikhef National Institute for Subatomic Physics and University of Amsterdam, Amsterdam, Netherlands\\
% ...
$^{115}$ Homer L. Dodge Department of Physics and Astronomy, University of Oklahoma, Norman OK, United States of America\\
% ...
$^{119}$ LAL, Univ. Paris-Sud, CNRS/IN2P3, Universit{\'e} Paris-Saclay, Orsay, France\\
% ...
$^{137}$ $^{(a)}$ Facult{\'e} des Sciences Ain Chock, R{\'e}seau Universitaire de Physique des Hautes Energies - Universit{\'e} Hassan II, Casablanca; $^{(b)}$ Centre National de l'Energie des Sciences Techniques Nucleaires, Rabat; $^{(c)}$ Facult{\'e} des Sciences Semlalia, Universit{\'e} Cadi Ayyad, LPHEA-Marrakech; $^{(d)}$ Facult{\'e} des Sciences, Universit{\'e} Mohamed Premier and LPTPM, Oujda; $^{(e)}$ Facult{\'e} des sciences, Universit{\'e} Mohammed V, Rabat, Morocco\\
% ...
$^{139}$ Santa Cruz Institute for Particle Physics, University of California Santa Cruz, Santa Cruz CA, United States of America\\
% ...
$^{151}$ Department of Physics and Astronomy, University of Sussex, Brighton, United Kingdom\\
% ...
$^{155}$ Raymond and Beverly Sackler School of Physics and Astronomy, Tel Aviv University, Tel Aviv, Israel\\
% ...
$^{a}$ Also at Department of Physics, King's College London, London, United Kingdom\\
$^{b}$ Also at Institute of Physics, Azerbaijan Academy of Sciences, Baku, Azerbaijan\\
$^{c}$ Also at Novosibirsk State University, Novosibirsk, Russia\\
$^{d}$ Also at TRIUMF, Vancouver BC, Canada\\
$^{e}$ Also at Department of Physics {\&} Astronomy, University of Louisville, Louisville, KY, United States of America\\
% ...
\end{flushleft}

\end{document}